\documentclass[aps,groupedaddress,fleqn,twocolumn,longbibliography]{revtex4-2}
\usepackage{graphicx}
\usepackage{xcolor}
\usepackage{amsmath,amssymb}
\usepackage{float}
\usepackage{physics}
\usepackage{amsfonts}
\usepackage{verbatim}
\usepackage{flushend}
\usepackage{balance}
\usepackage{endnotes}
\usepackage{footnote}
\usepackage{adjustbox}
\usepackage{gensymb}
\usepackage{subfigure}
\usepackage{float}
\usepackage{epigraph}
\usepackage{xcolor}
\usepackage[utf8]{inputenc}
\usepackage{setspace}
\newcommand{\beq}{\begin{eqnarray}}
\newcommand{\eeq}{\end{eqnarray}}
\newcommand{\kb}{k_{\mathrm{B}}}
\newcommand{\av}{\alpha_V}
\newcommand{\ath}{\alpha_T}
\newcommand{\ld}{\lambda_{\mathrm{d}}}
\newcommand{\tm}{\tau_{\mathrm{M}}}
\newcommand{\ct}{c_{\mathrm{T}}}

\usepackage{amsmath}

\usepackage[normalem]{ulem}

\begin{document}

\title{Thermodynamics and transport in molten chloride salts and their mixtures}

\author{ C. Cockrell$^{1,2,\ast}$, M.Withington $^{2}$, H. L. Devereux$^{2}$,  A. M. Elena$^{3}$, I. T. Todorov$^{3}$, Z. K. Liu${^4}$, S. L. Shang${^4}$, J. S. McCloy$^{5}$, P. A. Bingham${^6}$, K. Trachenko ${^2}$}
\address{$^1$ Nuclear Futures Institute, Bangor University, Bangor, LL57 1UT, UK}
\address{$^2$ School of Physical and Chemical Sciences, Queen Mary University of London, Mile End Road, London, E1 4NS, UK}
\address{$^3$ Scientific Computing Department, Science and Technology Facilities Council, Daresbury Laboratory, Keckwick Lane, Daresbury, WA4 4AD, UK}
\address{$^4$ Department of Materials Science and Engineering, The Pennsylvania State University, University Park, PA 16802, USA}
\address{$^5$ School of Mechanical and Materials Engineering, Washington State University, Pullman, WA, USA}
\address{$^6$ Materials and Engineering Research Institute, Sheffield Hallam University, Sheffield, S1 1WB, UK}
\address{\textcolor{black}{$\ast$ Corresponding author}}

\begin{abstract}
Molten salts are important in a number of energy applications, but the fundamental mechanisms operating in ionic liquids are poorly understood, particularly at higher temperatures. This is despite their candidacy for deployment in solar cells, next-generation nuclear reactors, and nuclear pyroprocessing. We perform extensive molecular dynamics simulations over a variety of molten chloride salt compositions at varying temperature and pressures to calculate the thermodynamic and transport properties of these liquids.  Using recent developments in the theory of liquid thermophysical properties, we interpret our results on the basis of collective atomistic dynamics (phonons). We find that the properties of ionic liquids well explained by their collective dynamics, as in simple liquids.  In particular, we relate the decrease of heat capacity, viscosity, and thermal conductivity to the loss of transverse phonons from the liquid spectrum. We observe the singular dependence of the isochoric heat capacity on the mean free path of phonons, and the obeyance of the Stokes-Einstein equation relating the viscosity to the mass diffusion. The transport properties of mixtures are more complicated compared to simple liquids, however viscosity and thermal conductivity are well guided by fundamental bounds proposed recently. The kinematic viscosity and thermal diffusivity lie very close to one another and obey the theoretical fundamental bounds determined solely by fundamental physical constants. Our results show that recent advances in the theoretical physics of liquids are applicable to molten salts mixtures, and therefore that the evolution and interplay of properties common to all liquids may act as a guide to a deeper understanding of these mixtures.
\end{abstract}

\maketitle

\section{Introduction}

High temperature liquids like molten salts, though poorly understood from a theoretical standpoint, will be integral to concentrated solar power plants and advanced nuclear reactors, while also lowering carbon footprint of heat-intensive industries \cite{Zhao2023}. They are central components to many proposed sections of a future nuclear industry, including coolant fluids, fuel transport fluids, and pyroprocessing hosts \cite{Rosenthal1970,Merle2009,Locatelli2013,Salanne2008,Locatelli2013,Uozumi2021,Lee2011,Mitachi2022}. Molten salt reactors enjoy an inherent safety over other working fluids, and can be designed to allow easy loading and unloading of the core, reducing waste production \cite{Merle2009,Locatelli2013}. Furthermore, molten salt reactors with flexible power output \cite{Denbow2020} and molten salts used as thermal storage media complement renewable energy sources \cite{Carabello2021,Bhatnagar2022,Yang2010}. Converting proposals and prototypes into wide-scale components of a next-generation power grid requires a heretofore unavailable understanding of the thermal hydraulics. The thermophysical properties of the liquid state are generally poorly understood, a problem exacerbated by the absence of a general thermodynamic theory of the liquid state which can guide experimental studies \cite{Trachenko2016}. 

It is often the case that real systems are more complicated than models and are harder to understand. In the case of molten salts, there is limited theory available on pure systems \cite{Hansen2003}, and almost no guide from theory regarding mixtures. Key properties determining thermal hydraulic performance are heat capacity, viscosity, thermal conductivity, and mass diffusivity. Measuring these properties at the high temperatures at which alkali halides melt is non-trivial due to corrosion and volatility. Acquiring thermal conductivity data in particular is challenging due to the obfuscating effects of convection and radiation \cite{Zhao2023}. This results in measurements from different experiments differing by a factor of up to three. Viscosity measurements are less problematic, though a theoretical model of the viscosity of molten salt \textit{mixtures} is still challenging \cite{Zhao2023b}. There is no predictive picture of molten salt thermodynamics and transport properties that can guide experiments and industry \cite{Zhao2023}. We have recently proposed a very general interrelation between transport properties and thermodynamics \cite{Cockrell2022,Cockrell2024b}, based on the phonon theory of liquid thermodynamics \cite{Trachenko2016}. Here we use these advances to make the first attempt to understand the thermodynamic and transport properties of molten salt mixtures on the basis of collective modes. This gives a  microscopic physical underpinning of these key properties of molten salts.

In this study, we perform molecular dynamics (MD) simulations on pure chloride salts and mixtures thereof in order to calculate the key properties mentioned above, heat capacity, viscosity, thermal conductivity, and diffusion coefficients, as a function of temperature across different isobars and isochores.  We find that in pure molten salts and in their mixtures, the thermodynamics and transport properties share the universal interrelation seen in other liquids \cite{Cockrell2024b}, meaning that the phonon theory of liquid thermodynamics applies to molten salts as well as simple liquids. The  kinematic viscosity and thermal diffusivity observe the universal limit informed by fundamental physical constants \cite{Trachenko2021,Trachenko2021b,Withington2024}, providing another theoretical guide to these systems. The phonon theory of liquid thermodynamics \cite{Bolmatov2012} was recently adapted to the thermal conductivity of dense fluids \cite{Zhao2021}, which generates good predictions for a variety of bonding types. Our findings extend this reasoning to more recent developments in liquid theory, and imply that the same basic physical mechanisms operate in ionic liquids as in noble, molecular, and metallic liquids, and that our understanding of the latter systems is transferable to the former.  This gives us predictive power in understanding currently studied molten salts and designing new ones.

\section{Methods}

\subsection{Interatomic interactions}

We model five different compositions: LiCl, KCl, and a binary, ternary, and quaternary mixture. The binary mixture is a lithium and potassium chloride eutectic (LKE) with chemical composition 0.58(LiCl) 0.42(KCl) \cite{Korin1997}. The ternary mixture is 10\% NaCl and 90\% LKE, and the quaternary mixture is 10\% NaCl, 10\% CsCl, and 80\% LKE. We choose chloride salt mixtures in order to learn the dynamical changes that take place as the composition of simple salts become more complex, giving insight into the changes that take place during the lifetime of a molten salt reactor. Classical interatomic interactions are modelled using electrostatics plus the Born-Huggins-Mayer pair-potential:

\begin{equation}
    \label{eqn:bhm}
    \varphi_{ij}(r) = A_{ij} \exp\left(B(\sigma_{ij} - r)\right) - \frac{C_{ij}}{r^6} - \frac{D_{ij}}{r^6}.
\end{equation}
Where $i$ and $j$ index species pairs. The parameter $B$ is the same between all pairs in the system. Pair parameters $A, B, \sigma, C$, and $D$ for pure systems (one anion species and one cation species) follow the Fumi and Tosi rules \cite{Fumi1964,Sangster1976}. For binary and ternary compounds, we use the mixing rules introduced by Larsen \textit{et. al.} \cite{Larsen1973}. Cations and anions are assigned their formal charges: $z = \pm 1$. Parameters appearing in Eq. \ref{eqn:bhm} for each pair in each system we simulate are listed in Tabs. \ref{tab:lkeparams}-\ref{tab:csparams}

\begin{table*}[ht]
  \centering
  \begin{tabular}{ |c|c|c|c|c|c|c| } 
  \hline
Species $i$ & Species $j$ & $A_{ij}$ (eV) & $B$ (\AA$^{-1}$) &  $\sigma_{ij}$ (\AA) & $C_{ij}$ (eV $\cross$ \AA${^6}$) & $D_{ij}$ (eV $\cross$ \AA${^6}$)  \\
 \hline
    Li & Li  & 0.4420 & 2.9455 & 1.632 & 0.04557 & 0.01875 \\ 
    Li & K & 0.3428 & 2.9455 & 2.279 & 0.7590 & 0.53057 \\
    Li & Cl & 0.291 & 2.9455 & 2.401 & 1.248 & 1.498 \\
    K & K & 0.2637 & 2.9455 & 2.926 & 15.1681  & 14.9808 \\
    K & Cl & 0.2110 & 2.9455 & 3.048 & 29.9616 & 45.5666 \\
    Cl & Cl & 0.1582 & 2.9455 & 3.170 & 73.7566 & 148.195 \\
\hline
\end{tabular}
\caption{Parameters modelling pair interactions in the LKE eutectic derived in this work.}
\label{tab:lkeparams}
  \end{table*}

\begin{table*}[ht]
  \centering
  \begin{tabular}{ |c|c|c|c|c|c|c| } 
  \hline
Species $i$ & Species $j$ & $A_{ij}$ (eV) & $B$ (\AA$^{-1}$) &  $\sigma_{ij}$ (\AA) & $C_{ij}$ (eV $\cross$ \AA${^6}$) & $D_{ij}$ (eV $\cross$ \AA${^8}$)  \\
 \hline
Li & Li & 0.4220 &  2.96877 & 1.632 & 0.04557 & 0.01875 \\
Li & Na & 0.3428  &  2.96877 & 1.986 &  0.21285 & 0.09425 \\
Li & K  & 0.3428  &  2.96877 & 2.279 &  0.7590 & 0.53057 \\
Li & Cl & 0.2901 &  2.96877 & 2.401 & 1.248 & 1.498 \\
Na & Na & 0.2637 &  2.96877 & 2.340 & 1.0487 & 0.4994 \\
Na & K  & 0.2637  &  2.96877 & 2.633 &  3.9112 &  2.8626 \\
Na & Cl & 0.2110 &  2.96877 & 2.755 & 6.991 & 8.674 \\
K  & K  & 0.2637 &  2.96877 & 2.926 & 15.1681 & 14.9808 \\
K  & Cl & 0.2110 &  2.96877 & 3.048 & 29.9616 & 45.567 \\
Cl & Cl & 0.1582 &  2.96877 & 3.170 & 73.6273 & 147.918 \\
\hline
\end{tabular}
\caption{Parameters modelling pair interactions in the LKE eutectic with 10\% NaCl derived in this work.}
\label{tab:naparams}
  \end{table*}

\begin{table*}[ht]
  \centering
  \begin{tabular}{ |c|c|c|c|c|c|c| } 
  \hline
Species $i$ & Species $j$ & $A_{ij}$ (eV) & $B$ (\AA$^{-1}$) &  $\sigma_{ij}$ (\AA) & $C_{ij}$ (eV $\cross$ \AA${^6}$) & $D_{ij}$ (eV $\cross$ \AA${^6}$)  \\
 \hline
Li &      Li      &     0.4220 &  2.9739 &   1.632 &    0.04557 &  0.01872 \\
Li &     Na      &     0.3428  & 2.9739 &  1.986 &  0.21285 & 0.09425 \\
Li &     K       &     0.3428  & 2.9739 &  2.279 &  0.7590 & 0.53057 \\
Li &     Cs      &     0.3428  & 2.9739 &  2.536 &  1.771  & 1.9831 \\
Li &     Cl      &     0.2901  & 2.9739 &  2.401 &  1.248 &  1.498 \\
Na &     Na      &     0.2637  & 2.9739 &  2.340 &  1.04867 & 0.4994 \\
Na &     K       &     0.2637  & 2.9739 &  2.633 &  3.9112 & 2.8626 \\
Na &     Cs      &     0.2637  & 2.9739 &  2.890 &  9.402 &  10.8367 \\
Na &     Cl      &     0.2110  & 2.9739 &  2.755 &  6.991 &  8.676 \\
K  &     K       &     0.2637  & 2.9739 &  2.926  & 15.17  & 14.98 \\
K  &     Cs      &     0.2637  & 2.9739 &  3.183 &  37.52  & 52.8373 \\
K  &     Cl      &     0.2110  & 2.9739 &  3.048 &   29.96  & 45.567 \\
Cs &     Cs      &     0.2637  & 2.9739 &  3.440 &  94.88 &  173.53 \\
Cs &     Cl      &     0.2110  & 2.9739 &  3.305 &  80.52 &  156.05 \\
Cl &     Cl      &     0.1582  & 2.9739 &  3.170  & 74.24  & 149.17 \\

\hline
\end{tabular}
\caption{Parameters modelling pair interactions in the LKE eutectic with 10\% NaCl and 10\% CsCl derived in this work.}
\label{tab:csparams}
  \end{table*}

\subsection{Equilibration and production runs}  

We conduct all MD simulations using the DL\_POLY package \cite{Todorov2006}, with the electrostatic interactions having been evaluated using the smooth particle mesh Ewald method. Systems sizes between 800-4000 atoms were simulated, with all thermal and transport properties showing no appreciable dependence on system size, as has been previously noted in calculations of viscosity \cite{Celebi2020,Kim2019,Cockrell2021}. Each system is initialised in a high temperature melt and equilibrated at the target temperature and pressure for 0.2 ns in the NPT ensemble using Nosé-Hoover thermostats and barostats, each with relaxation times of 1.0 ps. These equilibrated configurations are then reseeded with 80 different velocity distributions which run for 50 ps. \textcolor{black}{After these 50 ps, the uncorrelated initial velocity distributions have forward-evolved the system to create 80 configurations whose atomic positions and velocities are uncorrelated between each other}. These configurations are the starting configurations for the 0.5 ns production runs which were conducted in the NPT ensemble and the NVE ensemble. The timestep was 1 fs for all simulations.


\subsection{Calculation of thermodynamic and transport properties}

Transport properties are calculated using the Green-Kubo method \cite{Allen1991} - integrating time correlation functions of thermodynamic fluxes. The dynamic shear viscosity $\eta$ is calculated from the stress tensor autocorrelation function:

\begin{equation}
    \label{eqn:gkviscosity}
    \eta = \frac{V}{\kb T} \int_{0}^{\infty} \dd t \left\langle \sigma_{xy}(t) \sigma_{xy}(0),\right\rangle
\end{equation}
with temperature $T$, volume $V$, and Boltzmann's constant $\kb$, and where $\sigma_{xy}$ is an off-diagonal element of the microscopic stress tensor:
\begin{equation}
    \label{eqn:stresstensor}
    \sigma_{\alpha \beta} = -\frac{1}{V} \sum_{\i = 1}^{N} \left( m v^{i}_{\alpha} v^{j}_{\beta} - \frac{1}{2} \sum_{i \neq j} (r^{j}_\alpha - r^{i}_{\alpha}) F^{ij}_{\beta}  \right).
\end{equation}
Here Greek letters index Cartesian components, $\mathbf{r}^{i}$ is the position vector of particle $i$, $\mathbf{v}^{i}$ is the velocity of particle $i$, and $\mathbf{F}^{ij}$ is the force impressed upon particle $i$ by particle $j$.

The thermal conductivity is likewise calculated using Green-Kubo integrals. However in multicomponent fluids, in contrast to the shear viscosity which depends only on the stress tensor, multiple microscopic fluxes must be calculated in order to arrive at the phenomenological thermal conductivity \cite{Armstrong2014}. The first current is the microscopic energy current density, $\mathbf{j}_e(t)$:
\begin{equation}
    \label{eqn:qcurrent}
    \mathbf{j}_e(t) = \frac{1}{V} \sum_{i=1}^N \left( u^i \mathbf{v}^i + \frac{1}{2} \sum_{j\neq i} \mathbf{F}^{ij} \cdot \mathbf{v}^i \ (\mathbf{r}^{j} - \mathbf{r}^{i}) \right),
\end{equation}
with $u_i$ the total (kinetic plus potential) energy of particle $i\in 1, 2, \ldots N$, $\mathbf{f}_{ij}$ the total force impressed upon particle $i$ by particle $j$, $\mathbf{r}_{ij}$ the inter particle separation vector between particles $i$ and $j$, and $\mathbf{r}_i$ is the position of particle $i$.

This current is then correlated with partial momentum densities $\mathbf{j}_I(t)$ for each species $I$ in the mixture:
\begin{equation}
    \label{eqn:partialmomentum}
    \mathbf{j}_I(t) = \frac{1}{V} \sum_{i\in I} m_i \mathbf{v}_i(t).
\end{equation}
We then define kinetic coefficients $L_{e,e}$, $L_{e,I}$, and $L_{I,J}$ with Green-Kubo formulae:
\begin{equation}
    \label{eqn:LeeGK}
    L_{e,e} = \frac{V}{3 \kb} \int_0^\infty \dd t \ \langle \mathbf{j}_e(0) \cdot \mathbf{j}_e(t)\rangle,
\end{equation}
\textcolor{black}{\begin{equation}
    \label{eqn:LqIGK}
    L_{e,I} = \frac{V}{3 \kb} \int_0^\infty \dd t \ \langle \mathbf{j}_e(0) \cdot \mathbf{j}_{I}(t)\rangle,
\end{equation}}
\begin{equation}
    \label{eqn:LIJGK}
    L_{I,J} = \frac{V}{3 \kb} \int_0^\infty \dd t \ \langle \mathbf{j}_{I}(0) \cdot \mathbf{j}_{J}(t)\rangle.
\end{equation}
The resulting thermal conductivity $\lambda$ is then given by
\begin{equation}
    \label{eqn:heatfluxcross}
    \lambda = -\frac{1}{T^2}\left(L_{e,e} - \sum_{I=1}^{M-1} \sum_{J=1}^{M-1} L_{e,I} L^{-1}_{I,J} L_{e,J} \right),
\end{equation}
where $L^{-1}_{I,J}$ is the $I, J$ element of the inverse of the symmetric matrix whose elements are $L_{I,J}$. The kinetic coefficients involving the partial momentum densities are correction terms which ensure that $\lambda$ characterises the heat flow in response to a temperature gradient with all advective effect (including partial momentum densities) vanish \cite{deGroot1984}.

In addition to viscosity and thermal conductivity, we calculate normalised ``diffuse" forms of these quantities. The kinematic shear viscosity, $\nu$, quantifies the rate of velocity (rather than momentum) diffusion through the system, with definition:
\begin{equation}
    \label{eqn:kinematic}
    \nu = \frac{\eta}{\rho},
\end{equation}
with $\rho$ the mass density. Likewise, the thermal diffusivity $\ath$ quantifies the rate of temperature (rather than heat) diffusion through the system, with definition:
\begin{equation}
    \label{eqn:diffusivity}
    \ath = \frac{\lambda}{\tilde{c}_P},
\end{equation}
with $\tilde{c}_P$ the isobaric heat capacity per unit volume. These quantities have the same dimensions as the diffusion coefficient and likewise relate the spatial and temporal evolution of thermal forces.

The isobaric heat capacity, $C_P$, can be calculated directly from a molecular dynamics trajectory in the NPT-ensemble using fluctuations in enthalpy \cite{Stroker2021}:
\begin{equation}
    \label{eqn:cp}
    C_P = \frac{\langle H^2 \rangle - \langle H \rangle ^2}{\kb T^2}.
\end{equation}
Fluctuations in enthalpy in the NPT ensemble (and fluctuations in energy in the NVT ensemble) converge slowly in molecular dynamics simulations, and it is more economical in terms of computational cost to use indirect expressions. In this case, we use the relationship between the isochoric heat capacity $C_V$ and the deriatives of the Gibbs potential \cite{Callen1985}:
\begin{equation}
    \label{eqn:heatcapacityrelation}
    C_P = C_V + V T \av^2 B,
\end{equation}
where $\av$ is the isobaric coefficient of thermal expansion and $B$ is the isothermal bulk modulus. These quantaties are also calculated from fluctuations of extensive properties \cite{Stroker2021}:

\begin{equation}
    \label{eqn:alpha}
    \av = \frac{1}{\kb T^2} \frac{\langle H V \rangle - \langle H \rangle \langle V \rangle}{\langle V \rangle},
\end{equation}
and
\begin{equation}
    \label{eqn:bulk}
    B = \frac{\langle V \rangle \kb T}{\langle V^2 \rangle - \langle V \rangle^2},
\end{equation}
Of these, $\av$ is the slowest to converge, however it does so more quickly than $C_P$. On the other hand, $C_V$ can be calculated in the NVE ensemble using fluctuations in kinetic energy, $K$ \cite{Tuckerman2010}:
\begin{equation}
    \label{eqn:cv}
    C_V = \frac{3 N \kb}{2} \left( 1 - \frac{\langle K^2 \rangle - \langle K \rangle^2}{\frac{3}{2} N \kb^2 \langle T \rangle^2} \right)^{-1}.
\end{equation}
This expression also converges more quickly than $C_P$. We therefore use Eqs. \ref{eqn:heatcapacityrelation}-\ref{eqn:cv} to calculate $C_P$, though care must be taken that the mean density and temperature of the NPT and NVE simulations coincide.

Each quantity is calculated from each of the 80 uncorrelated trajectories at each of the target temperatures and densities. Quantities presented here are averaged over these trajectories.

\subsection{Interrelation between dynamics and thermodynamics}

It has previously \cite{Khrapak2024,Cockrell2024b,Cockrell2021,Cockrell2022,Dyre2018,Khrapak2021b} been demonstrated that the thermodynamics and transport properties of many liquids and supercritical fluids possess a unique and universal interrelation. It is possible to draw a one-to-one functional relationship between many of these properties. For example, a reduced viscosity is a single function of the excess entropy across the liquid phase diagram of many Lennard-Jones systems \cite{Khrapak2022c}, though it is expected that fluids with more complicated intermolecular interactions (departing from the simplest hard-sphere systems) will not observe this universality \cite{Heyes2023}. Here, we focus on the interrelation between the isochoric specific (per particle) heat capacity, $c_V$, and a quantity called the \textit{dynamic length}, $\ld$, which in liquids roughly corresponds to the mean free path of transverse collective modes (phonons) \cite{Cockrell2021}.

Particle dyamics in liquids combine periods of ``solid-like" oscillation within a locally elastic environment and diffusive ``jumps" from one environment to another. These jumps are infrequent enough for liquids to exhibit many solid-like properties over short timescales and lengthscales, including the ability to host transverse phonons. The jumps disrupt the elasticity and cause the rapid attenuation of transverse phonons. We estimate the mean free path using the Maxwell relaxation time, $\tm$ and the transverse speed of sound $\ct$. The former, introduced in the current context by Frenkel \cite{Frenkel1955}, approximates the timescale above and below which a liquid behaves viscously and elastically respectively, and has the following definition
\begin{equation}
    \label{eqn:maxwelltime}
    \tm = \frac{\eta}{G_{\infty}}.
\end{equation}
Here $G_{\infty}$ is the \textit{infinite frequency shear modulus}, characterising the elastic component of a system's response to shear stress in the infinite frequency limit. It is calculated using the zero-time of the correlation function used to calculate $\eta$ in Eq. \ref{eqn:gkviscosity}:
\begin{equation}
    \label{eqn:shearmod}
    G_{\infty} = \frac{V}{\kb T} \langle \sigma_{xy}^2 \rangle.
\end{equation}
This shear modulus is also used to define the transverse speed of sound in a fluid in a manner parallel to elasticity theory:
\begin{equation}
    \label{eqn:transversespeed}
    \ct = \sqrt{\frac{G_{\infty}}{\rho}}.
\end{equation}
It was then postulated that the product of these viscoelastic quantities approximates the maximum wavelength of transverse collective modes (phonons) propagating in the system (in the sense that a phonon whose wavelength exceeds the mean free path cannot travel even one wavelength before being scattered):
\begin{equation}
    \label{eqn:dynamiclength}
    \ld = \ct \tm.
\end{equation}
\textcolor{black}{System sizes ranged between around 26 \AA to 40 \AA for 800 atom systems, which exceeds the maximum $\ld$ calculated in this study. The dynamic length therefore represents a physically sensible lengthscale within the simulation cell.}

The relationship between $\ld$ and thermodynamics is made using the phonon theory of liquid thermodynamics \cite{Bolmatov2012,Trachenko2016,Cockrell2024b}. The reduction of heat capacity with temperature in the liquid state is explained by the decrease of $\ld$ with temperature (with $\tm$ being the predominant contributor). Increasing temperature increases atomic mobility, making jumps more frequent and reducing the timescales and lengthscales over which an atomic environment is elastic. This reduces the maximum wavelength of transverse phonons, which reduces the degrees of freedom which can express energy. To a first approximation the internal energy $U$ of the liquid depends on $\ld$ \textit{via}:
\begin{equation}
    \label{eqn:frenkelenergy}
    U = N \kb T \left(3 - \left(\frac{\tau_0}{\tau_{\mathrm{M}}}\right)^3\right),
\end{equation}
with $\tau_0$ a constant (in temperature). Multiplying the numerator and the denominator of the fraction gives $a_0/\ld$, with $a_0$ related to the interatomic length scale \cite{Trachenko2016}. It has been observed in MD simulations of many liquids and supercritical fluids with different bonding types \cite{Cockrell2024b} that the resultant heat capacity depends only on $\ld$, with all temperature and volume dependence of $c_V$ implicit in $\ld$ In other words, the heat capacity is entirely determined by $\ld$ along \textit{any} path on the phase diagram. Ionic liquids, however, have not yet been the object of this analysis.

\subsection{Fundamental bounds on liquid transport coefficients}

The general decrease of thermal conductivity and viscosity with increasing temperature in liquids and their general increase in gases implies the presence of a continuous minimum of each of these properties beyond the liquid-gas critical point. It turns out that there is a fundamental lower limit to kinematic viscosity $\nu$ and thermal diffusivity $\alpha_T$ in atomic condensed matter systems derived entirely from fundamental constants governing the formation of chemical bonds and interatomic forces. The minimal kinematic viscosity $\nu_{\mathrm{m}}$ can be approximated as \cite{Trachenko2020,Trachenko2021b}:

\begin{equation}
\nu_{\mathrm{m}}=\frac{1}{4\pi}\frac{\hbar}{\sqrt{m_e m}},
\label{eqn:etamin}
\end{equation}
where $m$ is the mass of the molecule and $m_e$ is the electron mass. The minimal thermal diffusivity, $\alpha_{\mathrm{m}}$, can be argued to take on the same value by similar considerations \cite{Trachenko2021b,Brazhkin2023}.

The smallest value the minima $\nu_{\mathrm{m}}$ and $\alpha_{\mathrm{m}}$ can assume in atomic condensed matter, with $m$ equal to a single proton mass, is of the order of 10$^{-7}$ $\frac{{\rm m}^2}{\rm s}$. This the value of this bound is indeed respected for noble and molecular liquids, water, and metallic alloys \cite{Trachenko2020, Gangopadhyay2022, Nussinov2022}. Eq. \ref{eqn:etamin} was derived under very general considerations, however the supercritical crossover between liquidlike and gaslike states \cite{Cockrell2021b} and the fundamental bounds on $\nu$ and $\alpha_T$ lying somewhere along this crossover, occurring at extreme temperatures, have not been experimentally investigated. Ionic liquids represent the last category of bonding type to which these analyses have not been applied.

\section{Results and discussion}

\subsection{Viscosity and diffusion}

\textcolor{black}{We now discuss the properties of each composition calculated along different phase diagram paths - we here consider isobars and isochores. We simulate up to 1800 K along isobars, as this is the point above which the simulated eutectic mixture boils - however we note that these rigid-ion potentials tend to fail to accurately predict the melting points, and presumably the boiling points, determined experimentally of the compositions they simulate.}

\begin{figure}
             \includegraphics[width=0.95\linewidth]{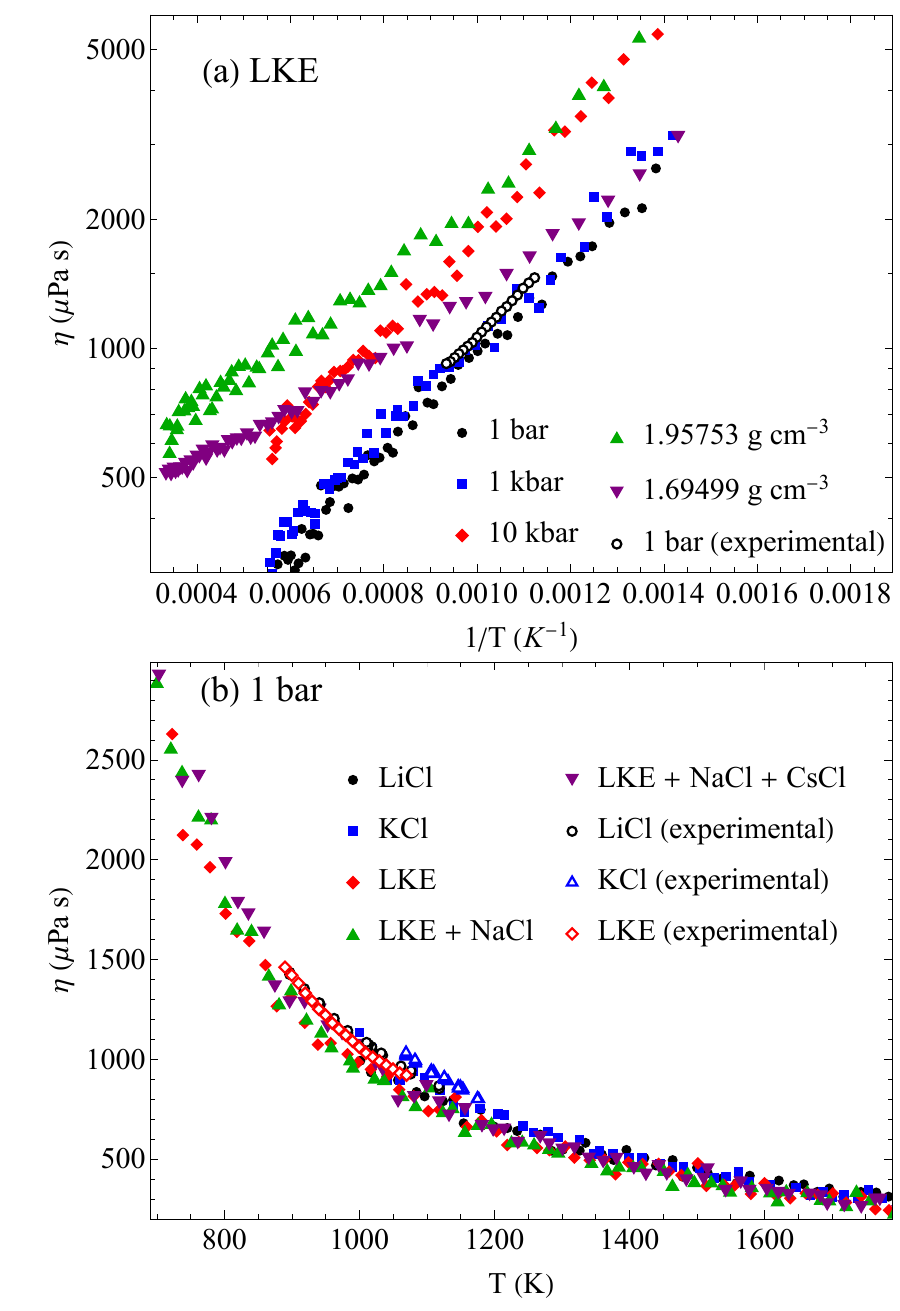}
    \caption{Shear viscosity $\eta$ of (a) LKE as a function of inverse temperature $1/T$ along different phase diagram paths; (b) LiCl, KCl, pure LKE, LKE with 10\% NaCl, and LKE with 10\% NaCl and 10\% CsCl as a function of $T$ at 1 bar. \textcolor{black}{Experimental data \cite{Brockner1975,Janz1975} are also included in these plots.}}
    \label{fig:etacolumn}
\end{figure}

We show the viscosity as a function of inverse temperature in each of our systems in Fig. \ref{fig:etacolumn}, noting that the 1 bar and 1 kbar isobars are mostly parallel and linear on the logarithmic axes, roughly indicating adherence to an Arrhenius-like equation for viscosity. The 10 kbar isobar and the two \textcolor{black}{isochores} slightly deviate from this linearity. Experimental viscosity of \textcolor{black}{the pure salts \cite{Brockner1975} and} the eutectic composition \cite{Janz1975}, also displayed on these axes, coincide very well with our simulated data. In Fig. \ref{fig:etacolumn}a we plot the viscosity of the LKE across 5 different phase diagram paths. In Fig. \ref{fig:etacolumn}b we compare our five different compositions at 1 bar. LiCl and KCl have similar viscosities in this temperature range. The eutectic mixture and high order mixtures exhibit very slightly lower viscosities than the pure salts, though this is hardly discernible beyond fluctuations. The difference in melting point, therefore, does not accompany a comparable difference in viscoelastic response. Though we do not plot these data here, we have also calculated the viscosity for mixtures with compositions 0.5(LiCl) 0.5(KCl) and 0.45(LiCl) 0.55(KCl), neither of which differs from the eutectic within the noise. \textcolor{black}{A relationship between ionic radii and viscosities of ionic liquids has been suggested before, and experimentally demonstrated in pure fluoride salts \cite{Popescu2015}. Such a simple relationship is not seen here, as LiCl and KCl possess nearly identical viscosities, and furthermore the viscosities (determined neither experimentally nor from simulations) of mixtures do not lie between those of the pure salts. This may indicate a highly collective character to the momentum transport in chlorides (compared with the fluorides in Ref. \cite{Popescu2015}) such that the properties of individual ions are less important. A full simulations study of fluoride mixtures would provide insight into this curious observation.}

\begin{figure}
             \includegraphics[width=0.95\linewidth]{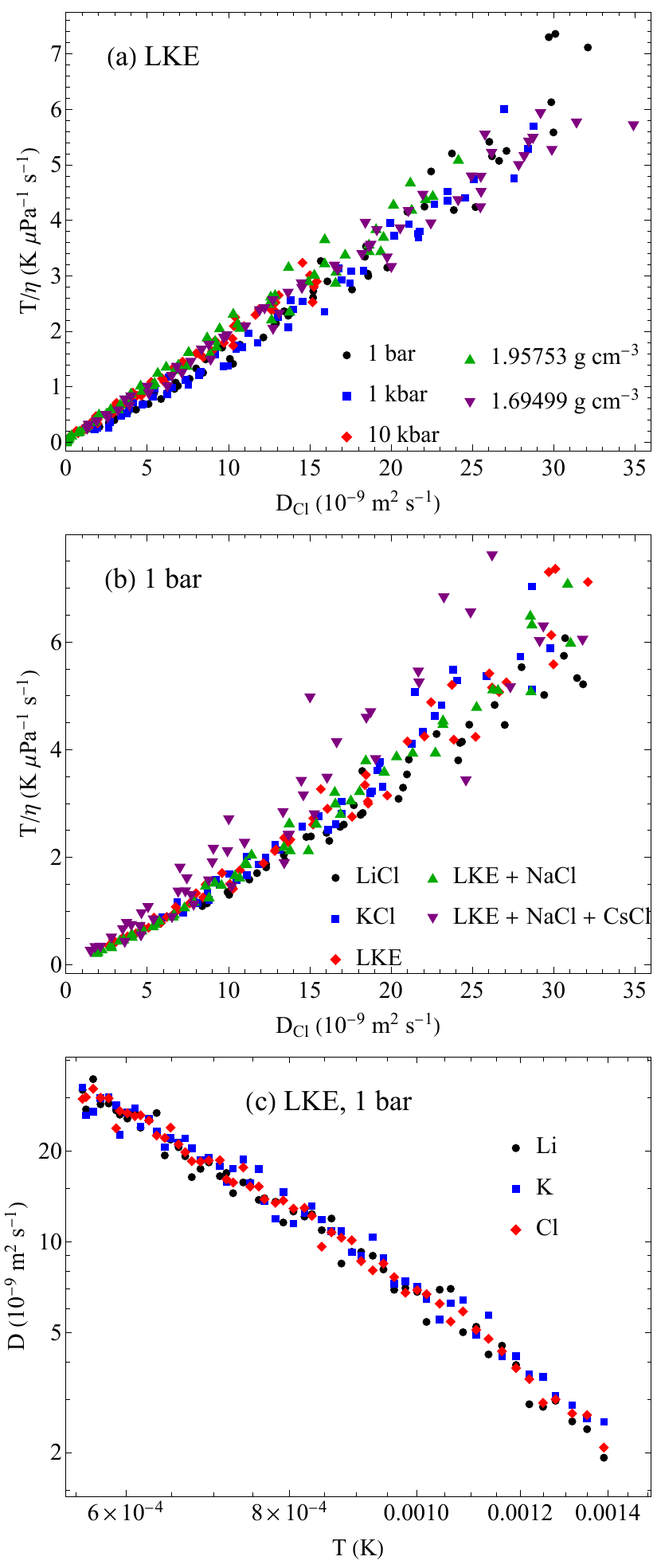}
    \caption{Relationship between the diffusion coefficient $D_{\mathrm{Cl}}$ of chlorine atoms and the quotient of the temperature $T$ by the shear viscosity $\eta$ in  (a) LKE along different phase diagram paths; (b) LiCl, KCl, pure LKE, and LKE with 10\% NaCl 1 bar. \textcolor{black}{(c) Diffusion coefficient $D$ of each species in LKE at 1 bar as a function of temperature.}}
    \label{fig:etadt}
\end{figure}

The Stokes-Einstein relation is a fluctuation-dissipation relation which connects the shear viscosity of a fluid with its self-diffusion coefficient, $D$. In fluids that observe this relation, $D$ may be expressed:
\begin{equation}
    \label{eqn:stokeseinstein}
    D = \frac{\kb T}{6 \pi r \eta}.
\end{equation}
In the original case of Brownian motion, $r$ is the radius of the Brownian particle. For the atomic self-diffusion, $r$ is an effective radius. We plot $T/\eta$ as a function of the diffusion coefficient of chlorine ions along several phase diagram paths in LKE in Fig. \ref{fig:etadt}a and along 1 bar in each composition in Fig. \ref{fig:etadt}b. \textcolor{black}{We plot the diffusion coefficient of chlorine because it is the only species common to all compositions, however we note that the diffusion coefficient of the cations are equal to those of chlorine in each mixture - we demonstrate this in LKE in Fig. \ref{fig:etadt}c. The lack of discrimination between the diffusivity of different species is due to the highly confined dynamics of the ions. Ions experience very little free motion which would allow a lower mass to be displaced further.}

A breakdown of the Stokes-Einstein relation implies that viscosity and atomic diffusion are determined by different timescales \cite{Kawasaki2017}, which may be brought about by the onset of highly correlated dynamics, spatially heterogeneous dynamics, such as in glass-forming supercooled liquids, or structural anomalies, such as in water \cite{Jeong2010,Chen2006,Kawasaki2017,Hodgdon1993}. Such effects may enhance the mobility of atoms but not larger-scale structural relaxations. Local atomic rearrangements have been observed to decorrelate from $\eta$ and $\tau_{\mathrm{M}}$ in ionic liquids \cite{Jeong2010}. The observance of the Stokes-Einstein relation here indicates that the liquid dynamics are ``simple" (such as those characterising liquid argon), with single-atom and larger structural relaxation occuring according to the same mechanism, governed by the Maxwell relaxation time $\tau_{\mathrm{M}}$.

\subsection{Thermodynamics}

\begin{figure}
             \includegraphics[width=0.95\linewidth]{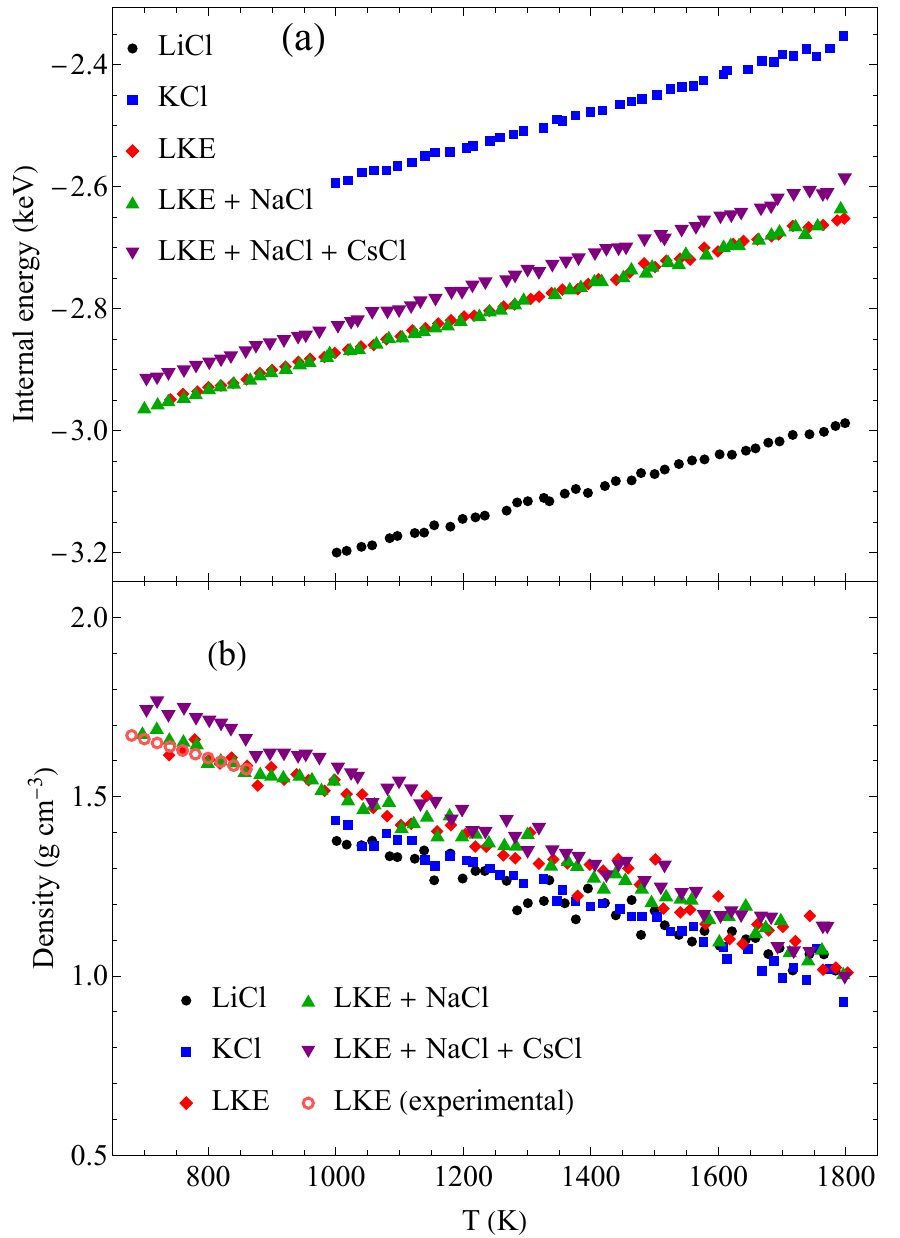}
    \caption{(a) Total internal (kinetic plus potential) energy of each composotion at 1 bar; (b) density of each composition at 1 bar\textcolor{black}{, including experimental measurements of the LKE \cite{Janz1975}}.}
    \label{fig:tdcolumn}
\end{figure}

We begin with elementary thermodynamic quantities in order to ascertain the basic impact of the mixture of pure salts. We plot the total internal (kinetic plus potential) energy and density in Fig. \ref{fig:tdcolumn}. LiCl has a considerably larger cohesive energy than KCl, with the mixtures lying between them. Notable the mixture with CsCl is somewhat less cohesive than the other two, whereas the addition of NaCl negligibly affects the total energy. Likewise, while the density of the pure salts, LKE, and LKE with NaCl are very similar, the addition of Cs considerably increases the density. Caesium is therefore less incorporated into the liquid structure, resulting in less cohesion and causing its large mass to increase the density. Available experimental measurements \cite{Janz1975} of the eutectic density over a limited temperature range are plot on these axes, and coincide exactly on our simulated data.

As described in Section 2, the heat capacity of many liquids has been shown to depend only on the dynamic length, $\ld$, which in liquids corresponds to the mean free path of transverse phonons. The heat capacity depends on temperature and density only \textit{via} $\ld$.

\begin{figure}
             \includegraphics[width=0.95\linewidth]{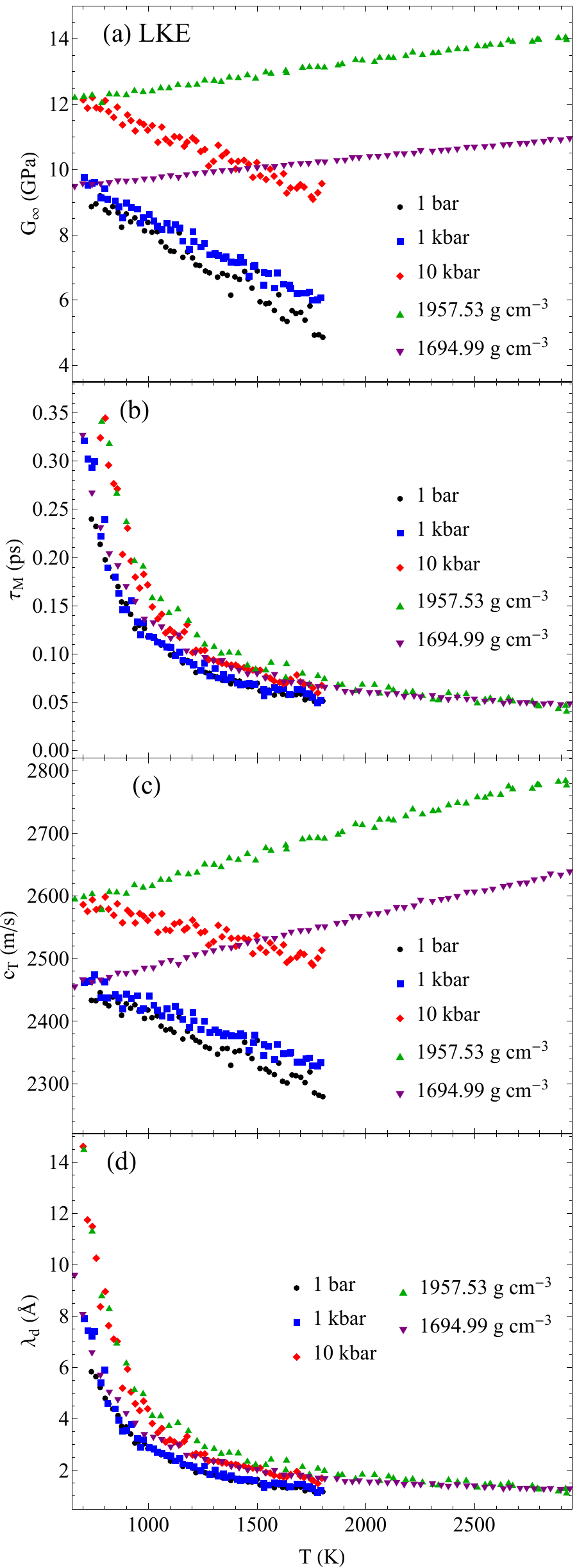}
    \caption{\textcolor{black}{(a) Maxwell relaxation time $\tm$; (b) infinite-frequency shear modulus $G_\infty$; and (c) transverse speed of sound $\ct$; and (d) dynamic length $\ld$ as a function of temperature across the phase diagram of LKE.} }
    \label{fig:cscolumn}
\end{figure}

\textcolor{black}{It is instructive first to look at the dynamic length $\ld$ and its constituent properties viscosity, $\eta$, the infinite-frequency shear modulus, $G_\infty$, and the density, $\rho$. We have discussed $\eta$ and $\rho$, and in Fig. \ref{fig:cscolumn}a-c we plot $G_\infty$, $\tm$, and $\ct$ as a function of temperature in LKE across different phase diagram paths (noting that the qualitative behaviour of these quantities in other compositions is identical).}

\textcolor{black}{Both $G_\infty$ and $\ct$ (which itself depends on $G_\infty$) decrease along isobars but increase along isochores, indicating that the negative contribution of increasing density on $G_\infty$ is greater in magnitude than the positive contribution of increasing temperature. The mixed response of $G_\infty$ along different phase diagram paths combine with the universal decrease of viscosity with increasing temperature (or decreasing pressure) to produce a relaxation time $\tm$ which also always decreases with increasing temperature (in the liquid state). Inspecting \ref{eqn:gkviscosity} we see that $\tm$ is effectively a normalised Green-Kubo integral - the viscosity divided by the autocorrelation function's initial value. The decrease of $\tm$ as $G_\infty$ increases indicates that the increase in elasticity represented by $G_\infty$ is purely ``kinetic". That is, the instantaneous response is larger due to a higher kinetic energy, however this high response is not sustained at lower frequencies and $\tm$ therefore decreases with increasing temperature.}

\textcolor{black}{The values of $\tm$ themselves are similar to those reported in liquid and supercritical argon \cite{Cockrell2021}, roughly 0.1 ps in magnitude. This decrease of $\tm$ is enough to compensate for the increase in $\ct$ along isochores, as the dynamic length $\ld$ also decreases with increasing temperature across all paths. This speed of sound $\ct$ is not a macroscopic speed of sound but rather the speed at which transverse waves would travel if the system's instantaneous response $G_\infty$ were maintained vanishing frequency. Its increase along isochores, like the increase of $G_\infty$, with increasing temperature is a purely kinetic effect, and the decreased lifetime (due to these fast kinetic correlations quickly being forgotten) of transverse excitations overwhelms this increasing speed, resulting in the decrease of $\ld$.}


\begin{figure}
             \includegraphics[width=0.95\linewidth]{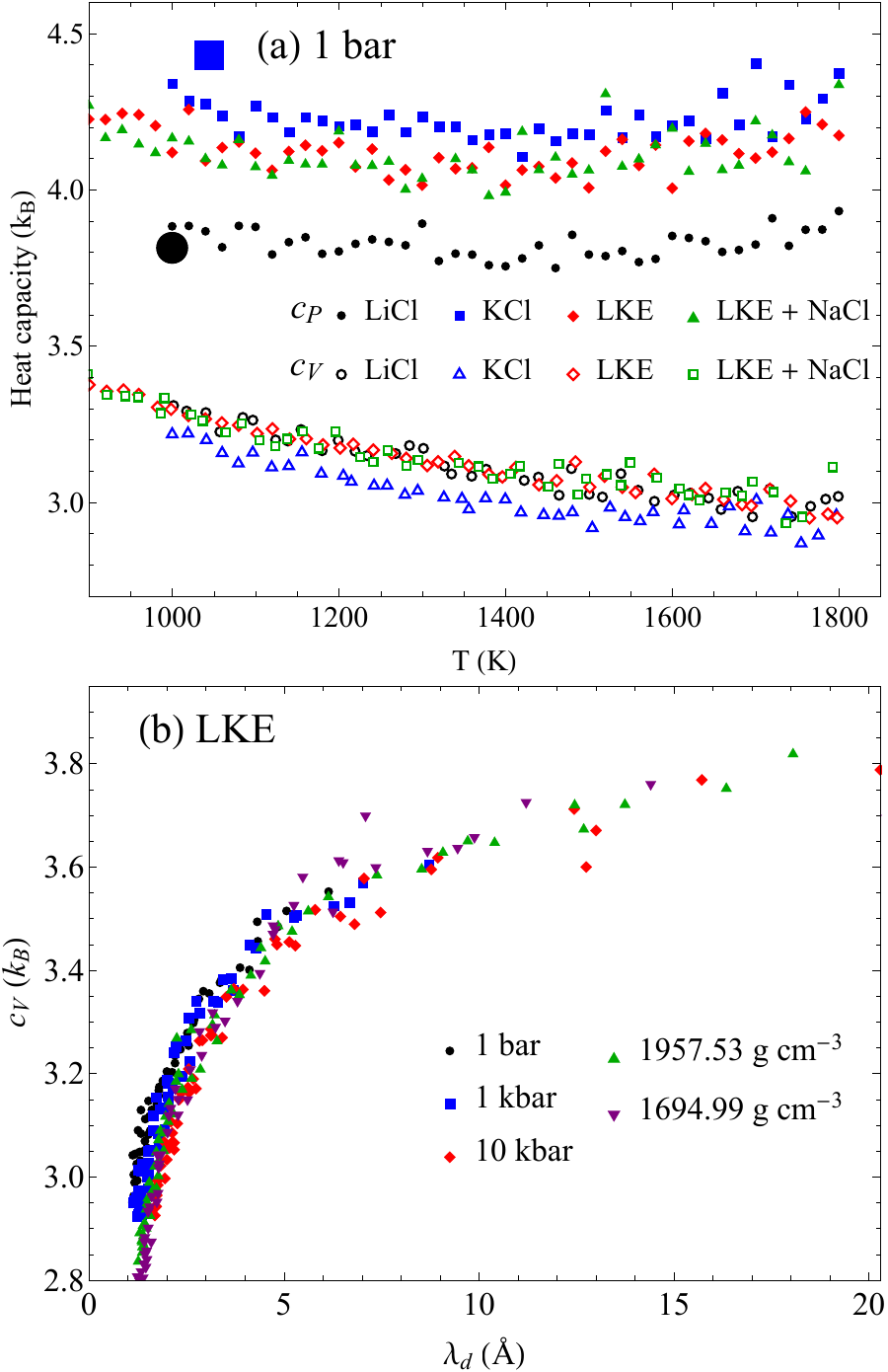}
    \caption{(a) Isochoric, $c_V$, and isobaric, $c_P$, heat capacities of the pure salts and the LKE at 1 bar. The large solid symbols are experimental data \cite{Nagasaka1992}; (b) isochoric heat capacity as a function of the dynamic length $\ld$ across the phase diagrams of LKE.}
    \label{fig:ccolumn}
\end{figure}

In Fig. \ref{fig:ccolumn}a we plot the isochoric and isobaric heat capacities of LiCl, KCl, LKE, and LKE with NaCl at 1 bar. The decrease of the isochoric heat capacity at higher temperatures coincides with the decrease of viscosity and loss of transverse collective modes as discussed above. The heat capacity of LiCl somewhat exceeds that of KCl, which can roughly be interpreted as the former having more degrees of freedom than the latter which can be satiated by temperature. This effect can also be brought about by anharmonic effects in the phonon spectrum \cite{Trachenko2016}. The $c_V$ of mixtures lie closer to that of LiCl than to that of KCl, implying that the introduction of KCl to LiCl does not significantly alter the highest-frequency modes (which are chiefly responsible for heat capacity), nor the introduction of NaCl to LKE. 

We observe that $c_V$ is close to its solid-like Dulong-Petit value. This is in agreement with Eq. \ref{eqn:frenkelenergy} showing that $c_V$ is close to 3 $\kb$ when $\tm \gg \tau_0$ and starts to deviate when $\tm$ becomes comparable to $\tau_0$, or when $\ld$ becomes comparable to the interatomic spacing. We note that on general grounds, it is possible to show that the liquid energy and heat capacity are determined, to a very good approximation, almost entirely by vibrational contributions across almost the entire range of $\tm$ \cite{Trachenko2013}. This assertion does not depend on a specific model of liquid energy and its dependence on the number of phonons.

The isobaric heat capacity $c_P$ is more complex, containing contributions of density derivatives as per Eq. \ref{eqn:heatcapacityrelation}. For $c_P$ the case is altered, with that of KCl comfortably exceeding that of LiCl. \textcolor{black}{This excession by KCl is primarily due to the larger thermal expansion coefficient of KCl than LiCl, and the thermal expansion of the mixtures are closer to that of KCl than LiCl. This results in $c_P$ of the mixtures lying much closer to that of KCl than LiCl}. These extra terms\textcolor{black}{, $\av$ and $B$,} weaken the correlation between microscopic dynamics and $c_P$ (compared to $c_V$) and introduce a correlation with longer range parameters related to the density.

\begin{figure}
             \includegraphics[width=0.95\linewidth]{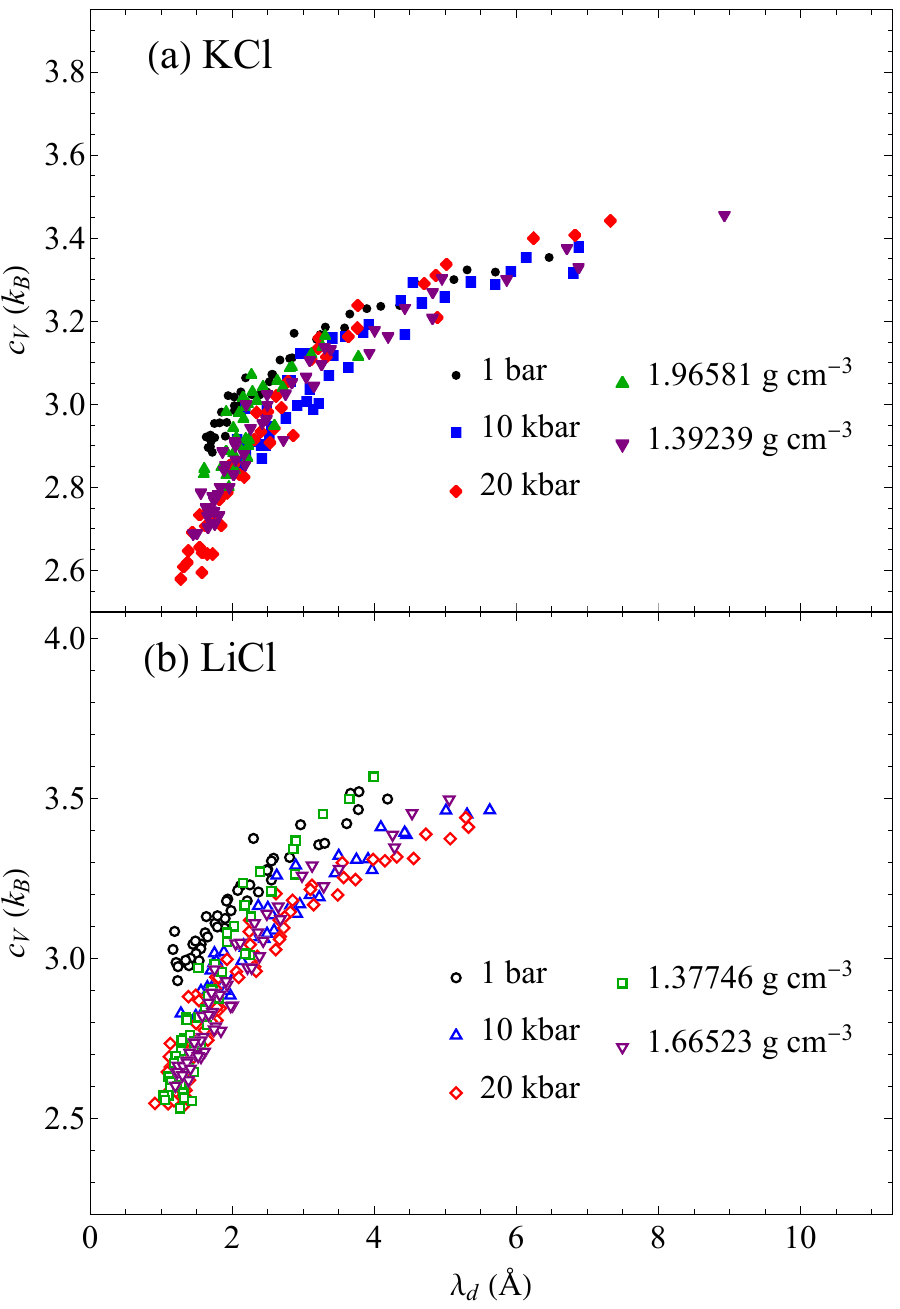}
    \caption{Isochoric heat capacity $c_V$ as a function of the dynamic length $\ld$ across the phase diagrams of (a) KCl (b) LiCl.}
    \label{fig:ccolumn2}
\end{figure}

In Fig. \ref{fig:ccolumn}b we display the isochoric heat capacity as a function of the dynamic length $\ld$ across five phase diagram paths. All curves collapse onto the ``main sequence" - the single curve characterising $c_V$ as a function of $\ld$ across the phase diagram. The same data are presented for KCl and LiCl in Fig. \ref{fig:ccolumn2}. In these compounds we see the same main sequence collapse, though LiCl at 1 bar falls slightly off of the main sequence. Ionic liquids therefore also observe the universal interrelation \cite{Cockrell2022} (called the ``c"-transition) and their thermodynamic properties are uniquely defined by the dynamic length, as was seen previously in noble, molecular, and metallic liquids \cite{Cockrell2021,Cockrell2024b,Cockrell2022}. This provides further indication that it is local directional bonds, such as those appearing in water and ammonia \cite{Cockrell2022,Cockrell2024b}, that cause a non-governance of heat capacity by collective modes and subsequent failure to observe the ``c"-transition (or similar thermodynamics-transport relations \cite{Fomin2010,Chopra2010,Abramson2007}. \textcolor{black}{The dynamic length $\ld$, interpreted as the maximum transverse phonon wavelength, decreases towards a minimum value of 1 \AA in LiCl, KCl, and the mixture, as it does in all other compounds in which it has been calculated \cite{Cockrell2022}. This is despite the difference in ionic radius between Li$^{+}$ and K$^{+}$. Indeed the minimum in $\ld$ in all compounds observing the ``c"-transition is roughly 1 \AA, indicating a very weak dependence of this quantity on the physical ``size" of the atoms and molecules participating in the collective modes.} Ionic liquids differ from other liquids investigated so far in the presence of multiple molecular species, which enables optical branches of the phonon spectrum. This may be related to the slight path dependence in the $c_V(\ld)$ curves of LiCl, as optical efects become more pronounced at larger mass ratios. We note that the path dependence on the main sequence, when speaking in terms of isobars, is maintained by the concomitant increase of $c_V$ and $\ld$ as pressure increases. The isochoric heat capacity of $c_V$ does not substantially differ as temperature increases. The relation between $c_V$ and $\ld$ is guaranteed if the former is chiefly governed by acoustic phonons. Optical transverse collective motion in liquids does not have a minimum wavevector like acoustic modes do (see, \textit{e.g.} \cite{Anento2001,Bryk2000}). This weakens the dependence of $c_V$ on $\ld$ and introduces the minor path dependence.

\subsection{Thermal conductivity}

The thermal conductivity measures the irreversible transport of heat, as shear viscosity measures that of shear momentum. Shear stress is decoupled from ``longitudinal" momentum fluctuations (those which propagate in the direction of the fluctuation), and the energy density is a ``longitudinal" property (like the bulk viscosity). Therefore, whereas diffusive particle motion is disruptive to transverse collective modes, diffusive motion alongside oscillatory motion constitutes a part of longitudinal collective modes (which is why gases can sustain longitudinal acoustic waves).

\begin{figure}
             \includegraphics[width=0.925\linewidth]{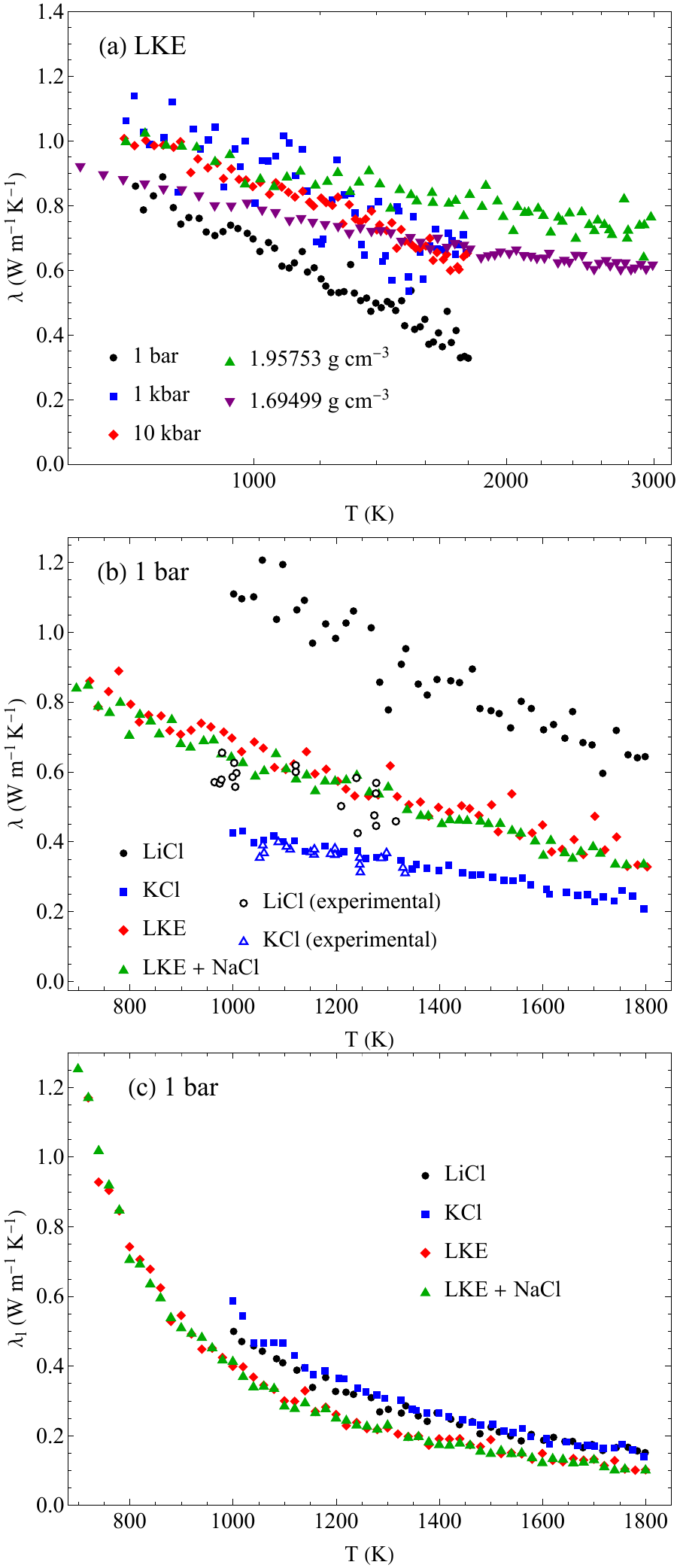}
    \caption{(a) Thermal conductivity $\lambda$ as a function of temperature $T$ in LKE along different phase diagram paths; \textcolor{black}{(b) $\lambda$ and (c) lattice thermal conductivity $\lambda_{\mathrm{l}}$ defined in Eq. \ref{eqn:latticethermal} as a function of LiCl, KCl, LKE, and LKE with 10\% NaCl at 1 bar.}}
    \label{fig:kappacolumn}
\end{figure}

The thermal conductivity of LKE along each phase diagram path is displayed in Fig. \ref{fig:kappacolumn}a, as was done for viscosity in Fig. \ref{fig:etacolumn}. As with viscosity, thermal conductivity decreases with increasing temperature across all paths. We note that at low temperature the thermal conductivity, in contrast with viscosity, of different paths closely coincides.

The thermal conductivity of different each composition, as displayed in Fig. \ref{fig:kappacolumn}b and again in contrast with viscosity, differs notably from one another. \textcolor{black}{Experimental data are available for the pure salts \cite{Nagasaka1992} and are also displayed in Fig. \ref{fig:kappacolumn}b. Though LiCl has a much larger conductivity than KCl, we note that the experimental conductivity of the former is slightly less than half the magnitude of the simulated conductivity, whereas the latter matches experimental measurements very well. The poor estimation of transport properties is a known effect of these rigid-ion models of molten salts \cite{Withington2024}, and may imply an excess either in the mobility or of the specific energy of the ions. We further note that experimental measurements which separate the effect of heat advection from thermal conductivity are challenging \cite{Robertson2022}.} As with thermodynamic properties, the thermal conductivity of each mixtures lies between those of the pure systems, though closer to that of KCl. This implies that the introduction of heavy cations to LiCl has a disruptive effect on the dynamics granting LiCl its high conductivity. 

\textcolor{black}{The lattice thermal conductivity $\lambda_s$ of crystalline solids may be approximated using:}
\textcolor{black}{\begin{equation}
    \lambda_{\mathrm{l}} = \frac{1}{3}\tilde{c}_P c_{\mathrm{s}} l,
    \label{eqn:latticethermal}
\end{equation}}
\textcolor{black}{where $\tilde{c}_P$ is the isobaric heat capacity per unit volume, defined above, $c_{\mathrm{s}}$ is the speed of sound, and $l$ is the mean free path of lattice phonons. On the basis that the liquid energy and thermodynamics may be understood by a phonon decomposition, it is reasonable to look to Eq. \ref{eqn:latticethermal} for instruction on the thermal conductivity of liquids \cite{Zhao2021}. As the microscopic energy density is a longitudinal variable, it is carried by longitudinal as well as transverse phonons. In this case, the longitudinal speed of sound (calcualated using the bulk modulus $B$ rather than the shear modulus $G_\infty$) would have to be incorporated into the calculation of $c_{\mathrm{s}}$, though this does not qualitatively change the evolution with temperature or density. Likewise, we have calculated the mean free path with the shear modulus, though exact the correspondence of $\ld$ with the heat capacity of supercritical fluids and dense gases (as with the heat capacity of liquids) implies that $\ld$ at the very least is closely related to the mean free path of longitudinal phonons \cite{Cockrell2021,Cockrell2022}, as transverse phonons are absent from these states of matter. We are therefore motivated to examine $\lambda$ in relation with $G_\infty, \ct$, and $\ld$.}

\textcolor{black}{The thermal conductivity increases with increasing density, but unlike $\ct$ and like $\eta$, $\lambda$ decreases with increasing temperature along isochores. This observation is explained by the concomitant decrease of $\ld$ along these phase diagram paths. This is due to the decrease of $\tm$ - though the speed of high-frequency collective modes increases, their lifetime very much decreases resulting in an overall less effective transport of energy. This explanation equally applies to viscosity and its evolution along different paths, however unlike viscosity the thermal conductivity of different compositions varies considerably. In Fig. \ref{fig:kappacolumn}c we plot $\lambda_{\mathrm{l}}$ of each composition as a function of temperature, and note this quantity hardly differs between each composition. This is essentially the same analysis as conceived by Zhao \textit{et. al.} \cite{Zhao2021}. The qualitative evolution of $\lambda_{\mathrm{l}}$ matches the directly calculated $\lambda$, but this approximation fails to capture the difference in magnitude of $\lambda$ exhibited by different compositions. This indicates that $\tilde{c}_P$, $\ct$, and $\ld$ (with the latter dominating the evolution of $\lambda_{\mathrm{l}}$) are at least somewhat appropriate markers of the thermal conductivity of these liquids, and therefore that the mechanisms behind these quantities (the energy and scattering of collective modes) may inform our understanding of liquid thermal conductivity. However, the failure of this model to capture the difference in magnitude between compositions indicates that there are other factors that are important in the transport of heat in ionic liquids.}

%


\subsection{Kinematic viscosity and thermal diffusivity}

\begin{figure}
             \includegraphics[width=0.925\linewidth]{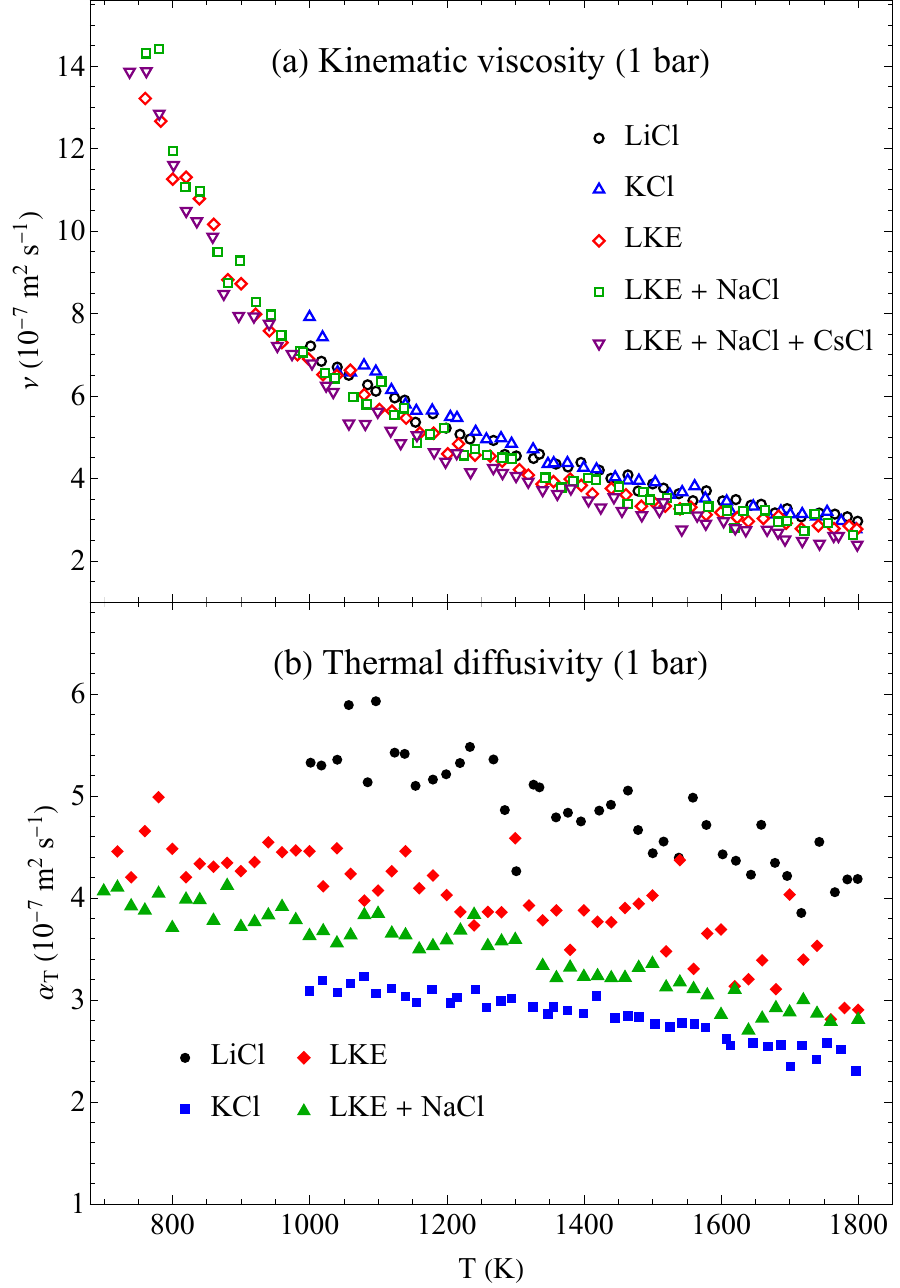}
    \caption{(a) Kinematic viscosity $\nu$; and thermal diffusivity $\alpha_{T}$ as a function of temperature at 1 bar in each composition.}
    \label{fig:bounds}
\end{figure}

Fig. \ref{fig:bounds} plots the kinematic viscosity and thermal diffusivity of each composition at 1 bar. As the viscosity and density hardly change between compositions, $\nu$ is also mostly insensitive to the changes in composition we study here. Heat capacity changes slightly more between compositions, however the plots of thermal diffusivity still qualitatively match those of thermal conductivity.

We here only investigate subcritical liquids, meaning that the minima in $\alpha_T$ and $\nu$, which occur beyond the critical point, are not observed - the boiling line intervenes before a minimum is reached. Nevertheless, the lower bound of around 10$^{-7}$ m$^2$s$^{-1}$ is observed by the thermal diffusivity and kinematic viscosity of all compositions. This observation of the fundamental minimum in viscosity plus the Stokes-Einstein relation also limits the maximum diffusion speed in these systems, as determined by Eq. \ref{eqn:stokeseinstein}. This implies that pure molten salts and their mixtures join the other liquid of different bonding types adhering to the minimum including noble, molecular, and hydrogen bonded liquids \cite{Trachenko2021,Brazhkin2023,Trachenko2021b}. 

\section{Conclusions}

We have investigated the thermodynamic and transport properties of chloride salts and their mixtures using molecular dynamics simulations. We interpret these properties in the light of recent developments in the understanding of liquid phonons and the fundamental bounds of transport properties in the liquid state. We find that the shear viscosity is mostly insensitive to the composition of chloride salt mixtures, despite the melting temperature of the eutectic (627 K) being far exceeded by the those of LiCl (878 K) and KCl (1043 K) \cite{Korin1997}. The viscosity and diffusion are interrelated by the Stokes-Einstein relation, as in other liquids. The thermodynamic interrelation between the dynamic length, $\ld$, determined from the viscosity, and the isochoric heat capacity $c_V$, is observed by each pure salt and the eutectic mixture. \textcolor{black}{The thermal conductivity is qualitatively well explained by a rudimentary phonon model, though the large conductivity of lithium-containing liquids is not adequately captured.} The diffusive transport properties, the thermal diffusivity and the kinematic viscosity, respect the fundamental bounds determined from fundamental constants. The combination of these observations suggest that the foundational mechanisms of the properties of ionic liquids such as these do not fundamentally differ from those of simpler liquids - indeed, ionic liquids are amenable to the phonon theory of liquid thermodynamics and associated theoretical models \cite{Bolmatov2012,Trachenko2016,Cockrell2024b,Yoon2018,Lin2003,Khrapak2024} and to the reasoning which determines the fundamental bounds of liquid transport properties \cite{Trachenko2021b}. The phonon theory of liquid thermodynamics \cite{Bolmatov2012} in its elementary form, designed at first to predict liquid heat capacity, has been adapted to produce a predictive model of liquid thermal conductivity \cite{Zhao2021}. The results we present here likewise imply that more recent developments \cite{Trachenko2016,Cockrell2024c,Yoon2018,Khrapak2024} can therefore likewise act as a spring board for a deeper and broader theoretical understanding of the thermodynamics and transport of ionic liquids and their relationships with each other and the microscopic dynamics.

\section*{Acknowledgements}
UK researchers are grateful to EPSRC (grant No. EP/X011607/1) and Queen Mary University of London for support. The U.S. researchers  acknowledge financial support by the U.S. Department of Energy  via Award No. DE-NE0009288. For all simulations conducted, this research utilised Queen Mary's Apocrita HPC facility, supported by QMUL Research-IT. http://doi.org/10.5281/zenodo.438045.

\bibliography{collection_2024}

\end{document}